\documentclass{sf2a-conf2017}
\usepackage{graphicx}
\usepackage{epstopdf}

\begin{document}

\TitreGlobal{SF2A 2017}


\title{Secular evolution of Milky Way-type galaxies}

\runningtitle{Secular evolution}

\author{F. Combes}\address{Observatoire de Paris, LERMA, College de France, CNRS, PSL, Sorbonne University UPMC, F-75014, Paris, France}

\setcounter{page}{237}


\maketitle


\begin{abstract}
The internal evolution of disk galaxies like the Milky Way
are driven by non-axisymmetries (bars) and the implied angular momentum transfer
of the matter; baryons are essentially driven inwards to build a more concentrated disk.
This mass concentration may lead to the decoupling of a secondary bar, since the 
orbit precessing frequency is then much enhanced. Vertical resonances with the bar will
form a box/peanut bulge in a Gyr time-scale.
Gas flows due to gravity torques can lead to a young nuclear disk forming stars,
revealed by a $\sigma$-drop in velocity dispersion. These gas flows moderated by feedback 
produce intermittent accretion of the super-massive black hole, and cycles of AGN activity.
The fountain effect due to nuclear star formation may lead to inclined, and even polar  nuclear disks.
\end{abstract}



\vspace{4mm}

How do we define secular evolution? This is slow and internal evolution
of a galaxy, which can be fueled by long-term gas accretion from cosmic filaments,
as opposed to violent evolution in galaxy mergers, or through interactions
with a galaxy cluster.
At least three types of galaxy formation and evolution can be considered: 
(i) the monolithic scenario, in which the gas collapses and forms stars
in a time shorter than the time-scale for clouds to collide and flatten into a disk.
This forms a spheroidal galaxy, or a bulge that can later accrete gas to form a disk.
(ii) the hierarchical scenario in which  the gas flattens in disks before forming stars,
leading to disky galaxies, and their interaction/merger with random angular momentum
lead to the formation of spheroids. (iii) the third possibility is just a branching of (ii)
in low-density environments, when mergers are rare, and the galaxy disks evolve
internally, and form boxy bulges from their own disk stars.

\section{Two kinds of bulges, classical and pseudo (box/peanut)}

Classical bulges are generally the result of major mergers,
with unaligned spins: the remnant is not flattened and has little rotation.
Its light profile has a Sersic index n=4 or higher. 
In minor mergers, disks are more easily conserved, while the classical bulge grows.

During secular evolution,  bars and vertical resonances elevate stars in the inner parts
into a pseudo-bulge: a component intermediate between a spheroid and a disk
(Combes \& Sanders 1981). It is flattened, rotating, and has an exponential light 
profile (n$\sim$1). They are more frequent in late-type galaxies.

Clumpy galaxies at high redshift can also form a classical bulge, 
through  dynamical friction of the massive clumps againt dark matter.
The formation of classical bulges is favored, and this makes even more 
difficult to form bulgeless galaxies. Observations tell us that the majority of
 late-type galaxies have no or little bulge today
(Kormendy \& Fisher 2008, Weinzirl et al 2009).

The fraction of pseudo-bulges has been quantified recently
by Fisher \& Drory (2016), as a function of stellar mass:
classical bulges begin to dominate only for stellar masses
larger than 5 10$^{10}$ M$_\odot$.
The impact of environment is important: there exists
half less pseudo-bulges in centrals with respect
to satellites and field galaxies (Mishra et al. 2017).

From HST images at high redshift (Goods-South) it was possible to decompose
galaxies in disks and bulges, and distinguish pseudo and classicals
(Sachdeva et al. 2017). Although pseudo bulges have masses about
half that of classicals, both bulges double in mass since z$\sim$1:
the mass fraction increases from 10 to 26\% for the pseudo,
and from 21 to 52\% for the classicals. This points towards 
secular evolution with at most minor mergers.

It might not be as easy to separate the formation of the two kinds
of bulges, since the dynamical evolution implies an angular momentum
transfer with the bar, ending with a
spin-up of the classical bulge (Saha et al. 2012). 
This is particularly important for low-mass classicals, but also
for higher masses (Saha et al. 2016).

\section{ Angular momentum transfer}
 
Cold disks form by transferring angular momentum (AM) outwards.
Bars, as negative AM waves, are amplified in the process.
Stars exchange AM only at resonances (unless the potential is varying).
The stars emit AM at ILR,  absorb at CR and OLR (Lynden-Bell \& Kalnajs 1972).
The AM is also absorbed by the dark matter halo (Athanassoula 2002).

During bar growth, more and more particles are trapped along the bar,
and their orbits are more elongated; the 
bar pattern speed $\Omega_b$ slows down, and the 
corotation  moves outwards, the bar is longer in the disk.

When gas is present in the disk, it feels the gravity torques from
the bar. It is driven inwards from CR, gives its AM to the bar, which
weakens the bar (Bournaud \& Combes 2002).
When disks are refilled with gas and become more massive, they can reform a bar
with higher  $\Omega_b$ (shorter bars).

\begin{figure}[ht!]
 \centering
 \includegraphics[width=0.8\textwidth,clip]{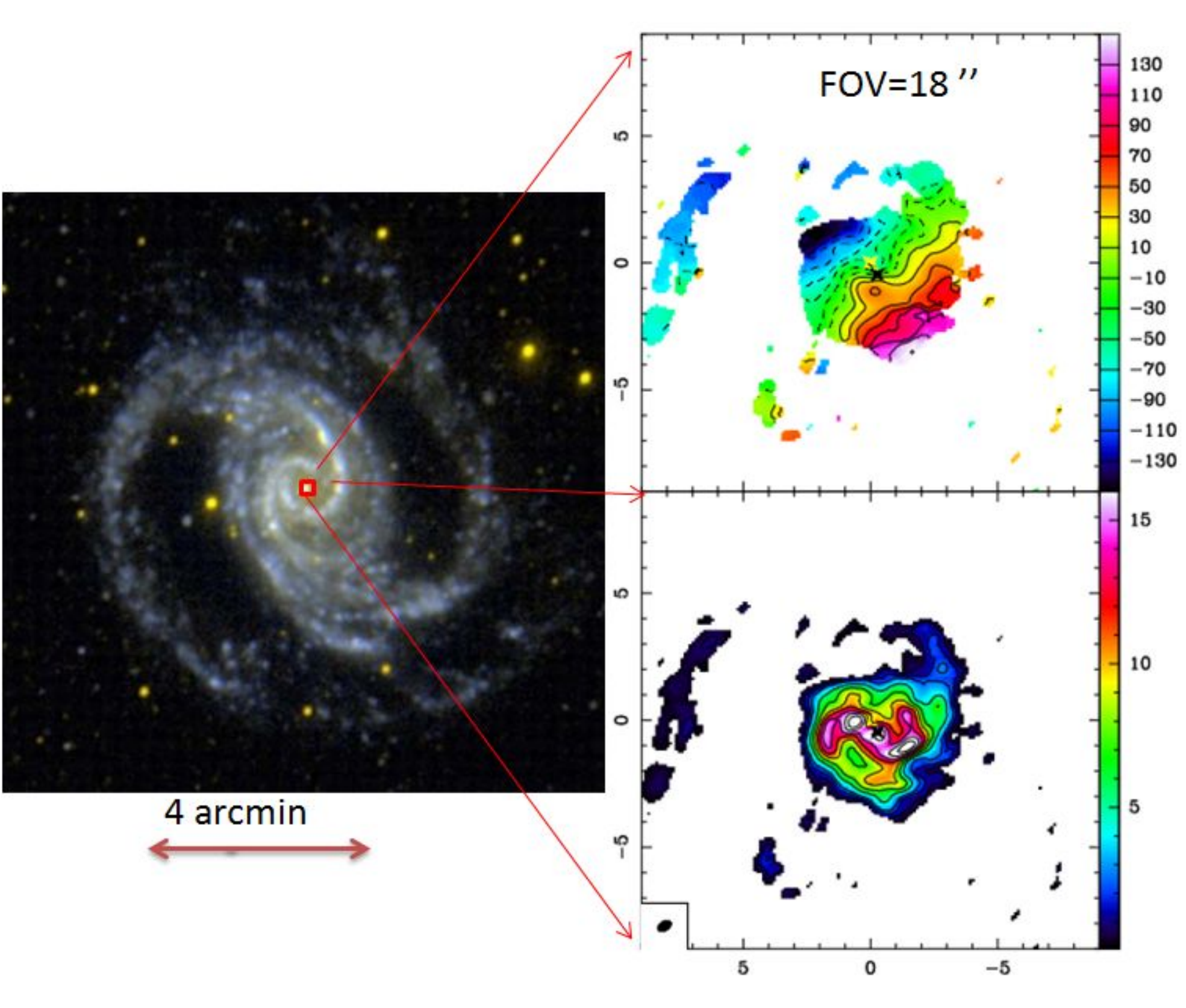}      
  \caption{ALMA CO(3-2) observations of NGC 1566 (Combes et al. 2014).
The field of view of the central molecular density and velocity map is 800 pc,
corresponding to the small red square in the right GALEX image. Note the conspicuous trailing 
nuclear spiral structure, causing the gas to fuel the Seyfert 1 nucleus.}
  \label{fig1}
\end{figure}

\subsection{Decoupling of a secondary bar}

When the bar has slown down and there exist two ILR,
the orbits become perpendicular (x2) to the bar in between the two ILR 
and do not sustain the bar anymore. This produces the decoupling of a
new faster bar inside the oILR ring. The z-resonance and formation of the peanut
also contribute to weaken the inner primary bar.
 The gravity torques of the second bar drive the gas to the center,
forming a cool nuclear disk with young stars, which might
be observed through a $\sigma$-drop, i.e. a dip in the stellar
velocity dispersion (Emsellem et al. 2001, Wozniak et al 2003).

New simulations of $\sigma$-drops have been done recently
(Portaluri et al. 2017, Di Matteo et al. 2017) including
spectral synthesis modelling, and chemical tagging.
They show clearly that the drop is seen in luminosity-weighted  images.

Embedded bars are observed in about 30\% of all barred galaxies.
It is also possible to form long-lived two-bar galaxies, with
no resonance in common,  no mode coupling
(Wozniak 2015). The star formation
in the gas stabilises the nuclear bar.
Also the nuclear bar could form first, in an
inside-out two bar formation scenario (Du et al. 2015),
in clumpy high-z galaxies.

\subsection{Bar gravity torques}

There exists a weak correlation between bars and 
AGN (Schawinski et al. 2010,  Cardamone et al. 2011)
and certainly bars help to drive gas to the nuclear region
by their gravity torques. But the situation is complex,
and depends on the various time-scales.

In a survey of 20 nearby Seyferts, we have been able
to compute the torques exerted by bars on the gas distribution,
obtaining the gravitational potential from HST
 red images (NUGA project).
At the scale of 10-100 pc, at which the
molecular gas maps were obtained, the statistics of fueling
are not high: only ~35\% of negative torques were measured in the center 
(e.g. Garcia-Burillo \& Combes 2012). The
rest of the times, the torques are positive, and
the gas is stalled in resonant rings. The
fueling phases are short, a few 10$^7$ yrs, may be
due to feedback. There is also
star formation fueled by the torques, always associated to AGN
activity, but with longer time-scales.

Embedded non-axisymmetries will occur at smaller-scales to control
gas accretion.
Zoomed simulations of gas accretion onto a central black hole
have revealed a cascade of m=2, m=1 perturbations
(Hopkins \& Quataert 2011),
providing an intermittent inflow rate. When the gas fraction is high,
the nuclear disk is unstable against warps, bending, and forms
clumps, sensitive to dynamical friction, which will drive gas inwards.

\begin{figure}[ht!]
 \centering
 \includegraphics[width=0.8\textwidth,clip]{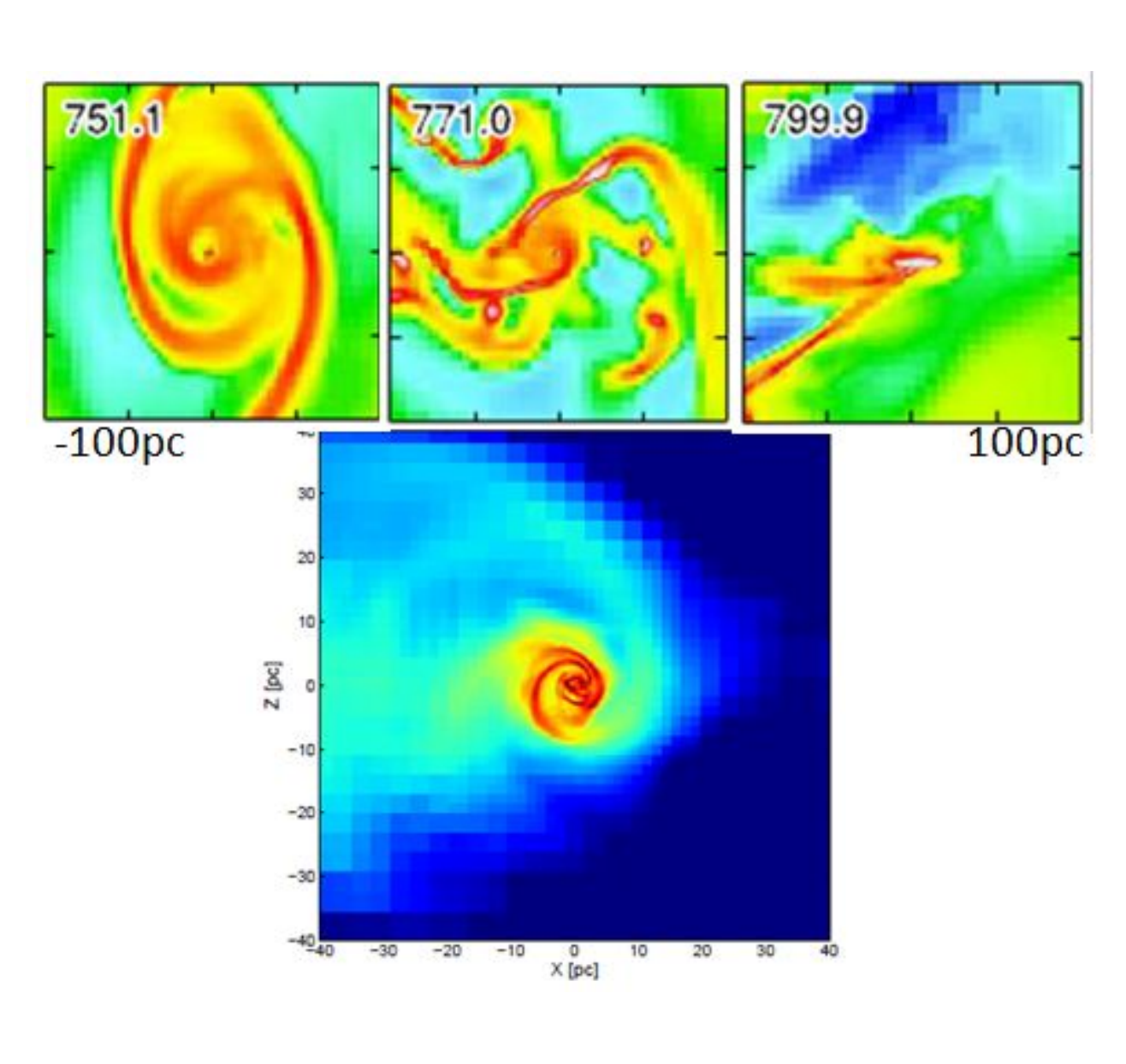}      
  \caption{{\bf Top:} Three snapshots of the gas in a MW-like galaxy simulation with AMR (from
 Renaud et al. 2016). This is the face-on view, but the last snapshot reveals a
perpendicular gas configuration (200 pc box). {\bf Bottom:} Edge-on view of the center (80 pc box),
where the gas now appears face-on, since it forms a polar nuclear disk around the black hole. }
  \label{fig2}
\end{figure}

High resolution (22pc) ALMA observations have been able to 
measure gravity torques even further towards the center.
In the barred spiral NGC 1433, a second resonant ring has been discovered
at 200 pc, at the ILR of the nuclear bar (Combes et al. 2013). But the torques are positive
inside, and the gas is piling up at this second ring (Smajic et al. 2014).
There is also a molecular outflow on the minor axis, 
due to the AGN feedback. This is a weak outflow of 100 km/s in velocity,
and dragging 7\% of the molecular gas mass.

The case of the nearby Seyfert type 1 galaxy NGC1566 is different:
trailing nuclear spiral arms have been discovered within 100pc around the black hole,
 torques are negative, fueling the nucleus (cf Figure \ref{fig1}).
This is due to the gravitational impact of the black hole, since the gas enters
its radius of influence.

In a high-resolution simulation meant to approach the MW, peculiar
hydrodynamical processes have been revealed in the central 200 pc region
(Renaud et al. 2016, Emsellem et al. 2015).
When the gas has sufficiently concentrated to the center, it becomes unstable,
fragments and forms stars. Through strong supernovae feedback and its fountain
effect, the gas is projected above the plane, and falls back to settle in a polar disk
around the black hole (cf Figure \ref{fig2}).

\section{Summary}

Secular evolution is the dominant scenario in Milky-Way type galaxies 
in the second part of the Universe age. The stellar bar favors the mass
concentration, and through vertical resonance, elevates stars 
to form a box/peanut bulge.
Primary bars drive gas from 10 kpc-scale  to R $\sim$100 pc, then
nuclear bars continue from 100 pc to 10 pc. Young nuclear disks are 
revealed by  $\sigma$-drops in their velocity dispersion.
 The mass concentration, and inward gas flows fuel nuclear starbursts and AGN.
At scales $\sim$1-10 pc, viscous turbulence, clumps, disk warps and bending,
take over to fuel the super-massive black hole.
The process is intermittent, moderated by
feedback and gas outflows.



%
\end{document}